\documentclass[showpacs,reprint,aps,prl]{revtex4-1}

\usepackage{isomath}
\usepackage{upgreek} % to have roman greek letter

\usepackage{epsfig, amsmath,amsfonts, amssymb,graphicx,amsthm,color}

\usepackage{mathtools}% For \MoveEqLeft
\usepackage{dcolumn}% Align table columns on decimal point
\usepackage{bm}% bold math
\usepackage{rotating}
\usepackage{amssymb}

\usepackage{graphicx}
\usepackage[utf8]{inputenc}

\usepackage{epstopdf}
\begin{document}

\title{Rovibrational optical cooling of a molecular beam}
\author{A. Cournol, P. Pillet, H. Lignier, D. Comparat}
\affiliation{Laboratoire Aim\'{e} Cotton, CNRS, Univ. Paris-Sud, ENS Paris Saclay, Universit\'e Paris-Saclay, B\^{a}t. 505, 91405 Orsay, France}

\date{\today}

\begin{abstract}
	Cooling the rotation and the vibration of molecules by broadband light sources was possible for trapped molecular ions or ultracold molecules. Because of a low power spectral density, the cooling timescale has never fell below than a few milliseconds.
Here we report on rotational and vibrational cooling of a supersonic beam of barium monofluoride molecules in less than 440 $\mu$s. Vibrational cooling was optimized by enhancing the spectral power density of a semiconductor light source at the underlying molecular transitions allowing us to transfer all the populations of $v''=1-3$ into the  vibrational ground state ($v''=0$).  Rotational cooling, that requires an efficient vibrational pumping, was then achieved. According to a Boltzmann fit, the rotation temperature was reduced by almost a factor of 10. In this fashion, the population of the lowest rotational levels increased by more than one order of magnitude.
\end{abstract}

\maketitle

Collisions, ionization, fragmentation, ablation or sputtering \cite{1992PhRvA..46..860K,1993PhyU...36..513K} are processes used for molecular formation in gas phase that are frequently implemented in laboratories. Molecules are thus formed in various distributions of internal and external degrees of freedom, the characteristics of which have a strong impact on experimental studies. In order to obtain better control of these molecular samples, some common techniques such as cryogenics and supersonic expansion are used to cool all degrees of freedom \cite{ramsey1985molecular,scoles1988,Pauly,pauly2013atom,lucas2013atomic,Morse1996}. However, these techniques cannot be triggered on demand, nor are they equally effective for all types of experiments \cite{sinha1973internal,mcclelland1979vibrational,zacharias1984rotational}. A method that, upon request, would effectively modify internal and external molecular states would be a breakthrough with multiple impacts in various domains of physics \cite{Carr2009,Hudson2011,Jin2012,Koch2016,Wall2016} and modern physical chemistry \cite{Shapiro2003,DAlessandro2007,Quemener2012,jankunas2015cold}. For example, such a tool, which would only act on internal states, should facilitate the decongestion of lines often encountered in spectroscopy, the preparation of states for collision studies, or even further optical manipulations (such as laser cooling of the molecular motion).\\
 %Also, molecular beams or samples with high density of particles per state could be produced by combining this technique to sympathetic or buffer gas cryogenic source \cite{hutzler2012a,2013Natur.495..490R,2014Natur.508...76H}.
In the last decade, it has become apparent that light can be useful for manipulating molecules \cite{Lemeshko2013}. Among other things, it was demonstrated that optical pumping of internal degrees of freedom is applicable to neutral molecules \cite{Viteau2008a,Manai2012a, Glockner2015,Hamamda2015a} and molecular ions \cite{Staanum2010,Schneider2010,Yzombard2015,Lien2014}. All the demonstrations were achieved with trapped or ultracold molecules because the cooling timescale ranged from a few ms to several seconds, which is far too long for free and fast particles. Experiments leading to the lowest observed timescale (ms) relied on broadband optical sources exciting many different molecular resonances.  In fact, the low power spectral density in the vicinity of these resonances implied a much longer timescale than that achieved by narrow linewidth lasers with common intensities ($\leq1$ W). Therefore, the development of a light source whose power density would be enhanced in the relevant spectral areas should accelerate optical cooling. In the present work, such an optimized light source was realized with a Tapered Amplifier (TA) and applied to supersonic beam of barium monofluoride (BaF). This allowed us to obtain a vibrational cooling, the speed of which made rotational cooling possible. 

Our experimental set-up is based on an ordinary seeded beam expanded from a supersonic nozzle with a 10 Hz repetition rate. Barium monofluoride (BaF) molecules were produced by chemical reaction between barium atoms ablated from a solid target by a pulsed laser and SF$_6$ molecules seeded in the carrier gas (Ar). Near the nozzle, collisions between BaF and Ar induced an effect of thermalization and the temperatures related to translation, rotation and vibration were $T_{\rm{trans}}\leq10$ K, $T_{\rm{rot}}\approx 50$ K and $T_{\rm{vib}}>500$ K respectively. The unequal rates of cooling for each degree of freedom is a well known fact that has already been reported \cite{Campbell2009,Barry2011}. This molecular beam penetrated a second room through a 3 mm skimmer where the local pressure was about $10^{-7}$ mbar, propagating freely over 25 cm for 440 $\mu$s and eventually passing through a 1.5 mm pinhole before detection.
During this time, three 1.5 W light beams with transverse dimensions smaller than $2$ mm were used for optical pumping: a diode laser at $\sim11170$ cm$^{-1}$ ($895$ nm) and a retro-injected optical TA at $\sim11760$ cm$^{-1}$ ($850$ nm) were dedicated to vibration whereas another diode laser at $\sim11630$ cm$^{-1}$ ($860$ nm) was used for rotation. They propagated in the opposite direction with respect to the molecular beam in order to maximize the interaction time. Their access to the vacuum chamber was enabled by mechanical shutters or Acousto-Optic Modulators (AOM) which allowed us to perform toggle experiments. Detection was ensured by Resonant Enhanced  Multi-Photon Ionization (REMPI) using a pulsed tunable laser based on a Optical Parametric Oscillator (OPO) at $\sim20400$ cm$^{-1}$ ($490$ nm) with 0.1 cm$^{-1}$ resolution and Micro Channel Plates (MCP) for subsequent photo-ionization. Scanning the OPO laser provided spectra of the C$^2 \Uppi_{1/2} (v'=v''+\Delta v,J')\leftarrow$X$^2 \Upsigma^+ (v'',J'',N'')$ transitions where primes refer to any excited state and seconds to the ground state; also $J', J''$ are total angular momenta and $N''$ is a rotational quantum number.
\begin{figure}
	\centerline{ \resizebox{1\linewidth}{!}{
			\includegraphics{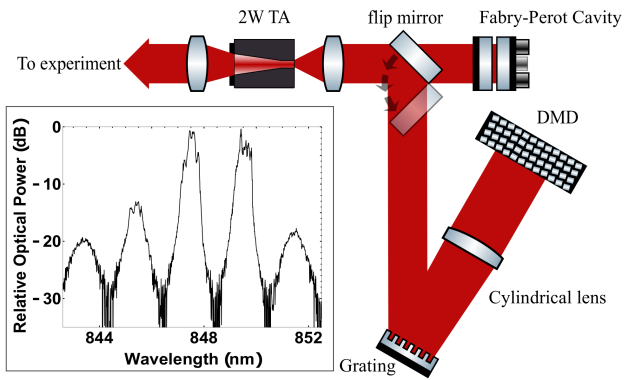}
		}}
		
		%    \centering
		%    \resizebox{0.95\linewidth}{!}
		%    {\includegraphics[1,1][523,345]{fig2}}
		\caption{Schematics of TA self-seeding for vibrational cooling laser. Two seeding optical lines can be selected by a flip mirror. The first line makes use of a FP cavity and the second line (4-f line) is based on a dispersion line consisting of a diffraction grating, a lens  and an array of micro-mirrors (cf. text for details). Inset: typical light spectrum obtained with the FP line used for vibrational cooling.
		}
		\label{schema_TA}
	\end{figure}

In first place, it is important to describe the principles of the vibrational pumping scheme that motivated the realization of a specific light source. The choice of optical transitions between the X and A electronic states was guided by the fact that leaks to other electronic states are negligible \cite{Chen2016}. Disregarding the rotation momentarily, the vibrational transition energies are given by $E(v', v'')=E(0,0)+\hbar \omega_{\rm{e}}' \Delta v+(\omega_{\rm{e}}'-\omega_{\rm{e}}'')v''$ within the framework of the harmonic approximation. For BaF, $\omega_{\rm{e}}''\approx\omega_{\rm{e}}'$ implies that the vibrational transitions are organized in sequences of constant values of $\Delta v$ roughly separated by $\hbar \omega_{\rm{e}}'$ i.e. $\sim450$ cm$^{-1}$. Also, in a given sequence $\Delta v$, the vibrational transitions are almost regularly separated by $\hbar(\omega_{\rm{e}}'-\omega_{\rm{e}}'')\approx 30$ cm$^{-1}$ giving rise to an excitation spectrum with a comb shape. 
A sideband cooling scheme can be implemented by using a light source covering the $\Delta v=-1$ transitions. In fact, it is unlikely that spontaneous emission will once again change the vibrational quantum number because the diagonal Franck-Condon factors $q_{v',v''}$ are very close to one for $v''=v'$. In other words, a single photon absorption has a high probability to reduce the vibrational quantum number by one unit. The particular structure of $q_{v',v''}$ in BaF is the origin of the simplicity of this vibrational cooling scheme. This advantage is compensated by small excitation strengths, given that $q_{v',v''}\ll 1$ when $v'\neq v''$. The consequence is that fast pumping requires important power spectral density in the vicinity of the molecular resonances.\\
A suitable light source was realized with the TA. The schematics used to tailor its spectrum is illustrated in Fig. \ref{schema_TA}. The idea consisted in injecting the TA by its own spontaneous emission (100 mW) submitted to a spectral shaping over $\sim130$ cm$^{-1}$ (10 nm) centered on $\sim11760$ cm$^{-1}$ (850 nm). Two commutable shaping lines were available. The first line was a Fabry-Perot (FP) cavity formed by two plane mirrors. The reflection on the cavity, used as a seeding, has a comb structure whose peaks are separated by the FP free spectral range and have a width determined by the reflectivity of the mirrors. By fixing the cavity length to $\sim 0.1$ mm, the TA spectrum had a few evenly-spaced peaks separated by $\sim 30$ cm$^{-1}$. The wavelength position could be finely adjusted by positioning one of the mirrors by a piezoelectric actuator. The stability of the cavity, checked with an optical spectrum analyzer (ANDO AQ6317B), was such that the peak drift was about 0.01 nm/h, slow enough to record data. As for the width of the peak, it was chosen so as to excite the whole rovibrational Q-branch. This required a width of $\sim2.7$ cm$^{-1}$ ($0.2$ nm), made possible by the use of 30\% reflective mirrors.
  The second shaping line was based on a dispersive grating and an array of micro-mirrors arranged in a folded 4-f line configuration as described in \cite{Viteau2008a}. It allowed us to select any tailored spectrum with $\sim0.7$ cm$^{-1}$ resolution. This grating-based shaping was more flexible and accurate than the FP cavity but gave access to a less extended spectrum owing to important power losses caused by the grating and the micro-mirrors.

\begin{figure}
	\centerline{ \resizebox{1\linewidth}{!}{
			\includegraphics{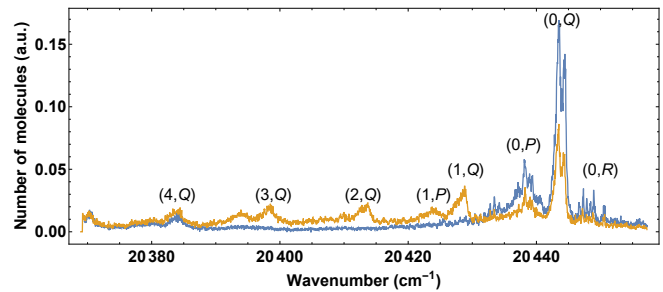}
		}}
		
		%    \centering
		%    \resizebox{0.95\linewidth}{!}
		%    {\includegraphics[1,1][523,345]{fig2}}
		\caption{Two REMPI spectra in the region of  (C$^2 \Uppi_{3/2},v'=v''+1,J')\leftarrow ( $X$^2 \Upsigma,v'',J''$) transitions. Peak labeling reads as $(v'',\Delta J'')$ where $\Delta J=J'-J''=-1, 0, +1$ are denoted by P, Q and R respectively. The orange spectrum is recorded without any optical pumping and the blue spectrum with optical vibration cooling. The corresponding light source is a TA with the typical spectrum inserted in Fig. \ref{schema_TA}. 
		}
		\label{vibrational_pumping}
	\end{figure}

The effect of the TA seeded by the FP line on BaF molecules was easily observed by recording two spectra in toggle mode as shown in Fig. \ref{vibrational_pumping}. It is clear that the populations of $v''=1-3$ were completely transferred into $v''=0$ whose population was multiplied by more than a factor 2. The transfer efficiency and the number of vibrational levels affected by the TA depends on the relative peak amplitudes of the light source. By tweaking a few geometrical features of the shaping line, the amplitudes were easily modified which impacted the results: it was observed that the population of much higher levels, up to $v''=7$, could be decreased but, in return, the efficiency to empty the lower levels , such as $v''=1,2$, fell apart rapidly. Two main causes can explain this observation. First, $q_{v''-1,v''}$ is a increasing function of $v''$  which means that the speed of optical pumping is reduced for low $v''$ values at a given optical power density. Then, the evolution of the population for a given $v''$ also depends on the inflow of population which lasts longer as $v''$ is small. A precise count of the number of molecules for each $v''$ was made difficult owing to several combined factors.  First of all, the REMPI signal for $v''>0$ was essentially limited to the Q-branches whose underlying rovibrational lines were unresolved by the OPO laser and consequently overlapped each other. This overlap and the resulting Q-branch amplitude was then sensitive to the OPO linewidth but also to a possible power broadening that depends on the FC factors of the X-C transitions and strongly varies with $v''$. 

However, the question of pumping efficiency was addressed quantitatively by using the TA with the 4-f line. This device also allowed us to produce a 1.5 W light source with a spectral distribution centered on $11797$ cm$^{-1}$ ($847.67$ nm) with  $\sim2.7$ cm$^{-1}$ ($0.2$ nm) linewidth, which corresponds to a vibrational pumping of the Q-branch of $v''=1$ only. Thanks to an AOM, the interaction time between the molecules and the light beam was scanned, providing a time-dependent depletion signal of $v''=1$. An exponential decay with a time constant of 30 $\mu$s was recorded, which is consistent with a simple rate model using experimental parameters without adjustment.

The fact that the vibrational pumping of $v''=1$ took much less time than the propagation time of the molecular beam paved the way for rotational cooling. Indeed, our rotational cooling scheme only involved molecules in $v''=0$ submitted to rovibrational cooling transitions A$^2 \Uppi_{1/2} (v'=0,J')\leftarrow$X$^2 \Upsigma^+ (v''=0, J'',N'')$. This type of transition is most appropriate because $q_{00}\sim0.95$, which typically implies that a dozen of absorption-spontaneous emission cycles could be induced before a change of vibrational state occurs. Because $q_{01}\gg q_{0v''}$ for $v''>1$, molecules escaping from $v''=0$ should reach $v''=1$ almost exclusively. 

Our rotational cooling scheme aimed at accumulating the maximum number of molecules in the lowest rotational levels. Necessarily, selection rules of optical transitions impose a maximum change $|\Delta J|\leq 1$ and a change of parity between the initial and final state. Given the spontaneous emission is out of control in a way, the lowering scheme relied on excitations causing $\Delta J=-1$ (called P-branch). In other words, the frequencies leading to $\Delta J=0,1$ (respectively Q and R-branches) had to be removed from the light spectrum. Considering the H\"{o}nl-London factors, an "absorption-emission" step driving only the P-branch lowers the $J$ value with a probability $3/4$ \cite{Herzberg1950}. Ideally, the succession of such steps should end up with two populated levels, each one corresponding to the lowest rotational level available for a given parity. For a more realistic description, it must be considered that the X and A states of BaF are described by two different Hund's cases. Instead of a simple P-branch, there are then $\rm{^PP}$, $\rm{^OP}$ transitions and also $\rm{^PQ}$ transitions (labeled as $^{\Delta N}\Delta J$) in the same spectral area \cite{Steimle2011}.

The broadband light source used for rotational cooling was designed to cover the spectral region occupied by these aforementioned transitions. A 4-f line, almost identical to that used with the TA, ensured a suitable spectral shaping by removing any light below $\sim11607$ cm$^{-1}$ and above $\sim11634$ cm$^{-1}$). The propagation factor of our diode ($M^2=21$ for the "best" axis) impacted the cut-off resolution that, considering the first order of diffraction and assuming Rayleigh criterion for the minimum resolvable detail, lead to a theoretical and measured resolution of $5.3$ cm$^{-1}$. This is 25 times larger than the X state rotational constant, $B_{v''=0}\approx 0.21$ cm$^{-1}$, i.e. the resolution needed to separate the lowest $J$ values of the P branch without affecting the Q and R-branches. 
To compensate the vibrational change occasioned by the rotational cooling, the TA source is used with its 4f-line to produce a 2 W light source with a 0.2 nm linewidth to induce Q-branch excitations of the  A$^2 \Uppi_{3/2} (v'=0)\leftarrow$X$^2 \Upsigma^+ (v''=1)$  transitions. As the spectral density turned out to be slightly insufficient to bring molecules back to X$^2\Upsigma (v''=0)$, the laser diode at $\sim11170$ cm$^{-1}$ was added with a spectral full half-width maximum of 2 nm to drive the A$^2 \Uppi_{1/2} (v'=0)\leftarrow$X$^2 \Upsigma^+ (v''=1)$ transitions. The difference of about 590 cm$^{-1}$ (45 nm) between the two similar fine transitions was sufficient to combine our light beams with a dichroic mirror. 
\begin{figure}
	\centerline{ \resizebox{1\linewidth}{!}{
			\includegraphics{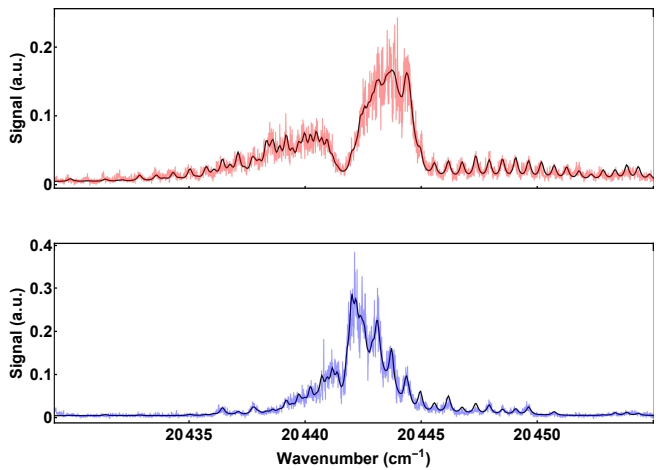}
		}}
		
		\caption{REMPI spectra recorded in toggle mode. Vibrational pumping $v''=0\leftarrow v''=1$ was always on, but, for the lower spectrum, the light source for rotational cooling was applied. The red and blue curves are the raw data and the black lines correspond to a fitting procedure. P, Q and R-branches are clearly separated in the upper spectrum because the rotational temperature is high enough whereas they are merged in the lower spectrum which indicates substantial rotational cooling.
		}

		\label{spectraRot}
	\end{figure}
	
	\begin{figure}
	\centerline{ \resizebox{0.9\linewidth}{!}{
			\includegraphics{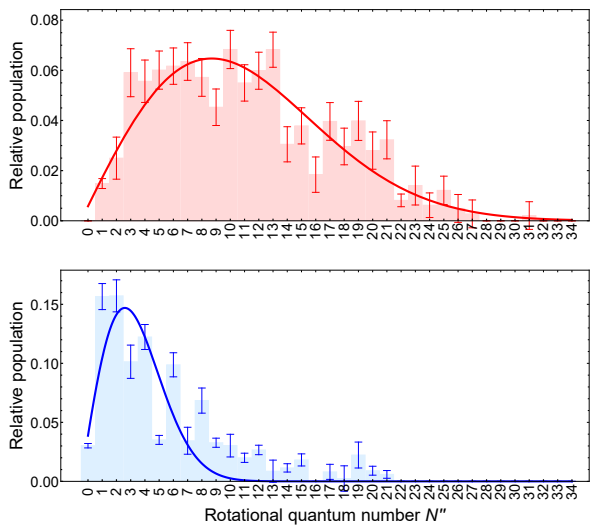}
		}}
		
		%    \centering
		%    \resizebox{0.95\linewidth}{!}
		%    {\includegraphics[1,1][523,345]{fig2}}
		\caption{Population distributions over the rotational quantum number $N''$ corresponding to the spectra in Fig. \ref{spectraRot}. The histograms and error bars result from the fitting procedure (see main text for details) whereas the solid lines correspond to fitted Maxwell-Boltzmann distributions of the histograms.
		}
		\label{popRot}
	\end{figure}

The effect of our rotational cooling scheme was tested by comparing REMPI spectra obtained in toggle mode.  A typical pair of unaveraged REMPI spectra is given in Fig. \ref{spectraRot}. The upper red spectrum reflects the rotational distribution of the X$^2 \Upsigma^+ (v''=0,J'',N'')$ with vibrational pumping only. This pumping transferred the initial population of $v''=1$ into $v''=0$ with presumably minor changes on the rotational distribution. On the other hand, the lower blue spectrum was obtained whith rotational cooling. The modification of the spectrum shape is attributed to a displacement of the maximum of the rotational distribution. This assertion was confirmed and quantified by fitting the spectra with a theoretical model where the populations of X$^2 \Upsigma^+ (v''=0,J'',N'')$ were the free paramaters. Although the REMPI resolution was insufficient to distinguish the lines of the P and Q-branches, the well separated lines from the R branch greatly facilitated our fitting procedure. The line positions were provided by molecular constants of the X$^2 \Upsigma^+ (v''=0,J'',N'')$ and C$^2 \Uppi_{1/2} (v'=1,J')$ state \cite{Steimle2011,Effantin1990a} while the transition intensities were calculated by PGopher \cite{Western2017}.The REMPI linewidth was calculated by taking into account the natural OPO linewidth and the power broadening. 
The fitting models are represented by the black lines in Fig. \ref{spectraRot} and the corresponding rotational resulting populations are represented in Fig. \ref{popRot}. The upper red (resp. lower blue) distribution corresponds to the rotational population of the spectrum without (resp. with) rotational cooling  whereas the red and blue lines are the best fitted Maxwell-Boltzman (MB) distributions. For the distribution without rotational cooling, the MB temperature and the mean energy are  $T_\mathrm{rot}= 53$ K and $\langle E_\mathrm{rot}\rangle/kB=52$ K respectively. For the distribution obtained with rotational cooling, the same calculated quantities are $T_\mathrm{rot}= 6$ K and $\langle  E_\mathrm{rot}\rangle/kB = 18$ K. The difference originates from the fact that rotational cooling is not a thermalization process. However, whatever the description, an unambiguous rotational cooling has been observed leading to multiply the population of some low rotational levels up to a factor 18 if the effects of rotational and vibrational cooling are combined (cf. Fig. \ref{popRot}).

In conclusion, prior works demonstrated that a light source with a broadband spectrum was able to cool the internal degrees of freedom. The light spectrum was then shaped by removing power in well chosen spectral regions. This approach was limited to trapped molecular ions or ultracold molecules because the photon absorption rate was too low for unconfined molecules. Here the use of a TA allowed us to tailor the light spectrum without power losses which had been a key element to cool the vibration of BaF molecules. Although the shape of the light source spectrum is simplified owing to the unusual structure of BaF, the method should be applicable to other species if the use of micro-mirrors generates the suitable spectrum. It is worth insisting on the fact that rotational cooling had been accessible because of the efficiency of vibrational cooling. A deeper rotational cooling should be accessible by improving the resolution of the spectral cut-off. This could be improved by a light source with a quality factor $M^2 \sim 1$. Some other limitations could arise from the existence of dark states given that the light is polarized and the decrease of $J''$ goes with a reduction of the number of sub-magnetic levels. Some preliminary experiments seem to have indicated that this effect was not relevant for the current experiment. 	

The authors thank T. Courageux for experimental assistance. This work was supported by ANR MolSisCool, Dim Nano-K CPMV, CEFIPRA No. 5404-1, the European Research Council under the grant agreement n. 277762 COLDNANO and 712718-LASFIB.

%\begin{figure}
%	\centerline{ \resizebox{1\linewidth}{!}{
%			\includegraphics{4flineAndSPectrum.eps}
%		}}
%		
%		%    \centering
%		%    \resizebox{0.95\linewidth}{!}
%		%    {\includegraphics[1,1][523,345]{fig2}}
%		\caption{Optical setup of the 4-f line for frequency shaping of the rotational cooling laser.
%		}
%		\label{4fline}
%	\end{figure}

%\bibliographystyle{h-physrev}

\bibliographystyle{apsrev4-1}

\begin{thebibliography}{36}%
\makeatletter
\providecommand \@ifxundefined [1]{%
 \@ifx{#1\undefined}
}%
\providecommand \@ifnum [1]{%
 \ifnum #1\expandafter \@firstoftwo
 \else \expandafter \@secondoftwo
 \fi
}%
\providecommand \@ifx [1]{%
 \ifx #1\expandafter \@firstoftwo
 \else \expandafter \@secondoftwo
 \fi
}%
\providecommand \natexlab [1]{#1}%
\providecommand \enquote  [1]{``#1''}%
\providecommand \bibnamefont  [1]{#1}%
\providecommand \bibfnamefont [1]{#1}%
\providecommand \citenamefont [1]{#1}%
\providecommand \href@noop [0]{\@secondoftwo}%
\providecommand \href [0]{\begingroup \@sanitize@url \@href}%
\providecommand \@href[1]{\@@startlink{#1}\@@href}%
\providecommand \@@href[1]{\endgroup#1\@@endlink}%
\providecommand \@sanitize@url [0]{\catcode `\\12\catcode `\$12\catcode
  `\&12\catcode `\#12\catcode `\^12\catcode `\_12\catcode `\%12\relax}%
\providecommand \@@startlink[1]{}%
\providecommand \@@endlink[0]{}%
\providecommand \url  [0]{\begingroup\@sanitize@url \@url }%
\providecommand \@url [1]{\endgroup\@href {#1}{\urlprefix }}%
\providecommand \urlprefix  [0]{URL }%
\providecommand \Eprint [0]{\href }%
\providecommand \doibase [0]{http://dx.doi.org/}%
\providecommand \selectlanguage [0]{\@gobble}%
\providecommand \bibinfo  [0]{\@secondoftwo}%
\providecommand \bibfield  [0]{\@secondoftwo}%
\providecommand \translation [1]{[#1]}%
\providecommand \BibitemOpen [0]{}%
\providecommand \bibitemStop [0]{}%
\providecommand \bibitemNoStop [0]{.\EOS\space}%
\providecommand \EOS [0]{\spacefactor3000\relax}%
\providecommand \BibitemShut  [1]{\csname bibitem#1\endcsname}%
\let\auto@bib@innerbib\@empty
%</preamble>
\bibitem [{\citenamefont {Kelly}(1992)}]{1992PhRvA..46..860K}%
  \BibitemOpen
  \bibfield  {author} {\bibinfo {author} {\bibfnamefont {R.}~\bibnamefont
  {Kelly}},\ }\href {\doibase 10.1103/PhysRevA.46.860} {\bibfield  {journal}
  {\bibinfo  {journal} {Phys. Rev. A}\ }\textbf {\bibinfo {volume} {46}},\
  \bibinfo {pages} {860} (\bibinfo {year} {1992})}\BibitemShut {NoStop}%
\bibitem [{\citenamefont {Kudryavtsev}\ \emph {et~al.}(1993)\citenamefont
  {Kudryavtsev}, \citenamefont {Mazyar},\ and\ \citenamefont
  {Sukhov}}]{1993PhyU...36..513K}%
  \BibitemOpen
  \bibfield  {author} {\bibinfo {author} {\bibfnamefont {N.~N.}\ \bibnamefont
  {Kudryavtsev}}, \bibinfo {author} {\bibfnamefont {O.~A.}\ \bibnamefont
  {Mazyar}}, \ and\ \bibinfo {author} {\bibfnamefont {A.~M.}\ \bibnamefont
  {Sukhov}},\ }\href {\doibase 10.1070/PU1993v036n06ABEH002164} {\bibfield
  {journal} {\bibinfo  {journal} {Phys. Uspekhi}\ }\textbf {\bibinfo {volume}
  {36}},\ \bibinfo {pages} {513} (\bibinfo {year} {1993})}\BibitemShut
  {NoStop}%
\bibitem [{\citenamefont {Ramsey}(1985)}]{ramsey1985molecular}%
  \BibitemOpen
  \bibfield  {author} {\bibinfo {author} {\bibfnamefont {N.}~\bibnamefont
  {Ramsey}},\ }\href@noop {} {\emph {\bibinfo {title} {{Molecular beams}}}}\
  (\bibinfo  {publisher} {Oxford University Press},\ \bibinfo {year}
  {1985})\BibitemShut {NoStop}%
\bibitem [{\citenamefont {Scoles}(1988)}]{scoles1988}%
  \BibitemOpen
  \bibfield  {author} {\bibinfo {author} {\bibfnamefont {G.}~\bibnamefont
  {Scoles}},\ }\href {https://books.google.com/books?id=AohMcgAACAAJ{\&}pgis=1}
  {\emph {\bibinfo {title} {{Atomic and Molecular Beam Methods. - V. 1
  (1988)}}}}\ (\bibinfo  {publisher} {Oxford University Press},\ \bibinfo
  {year} {1988})\BibitemShut {NoStop}%
\bibitem [{\citenamefont {Pauly}(2000)}]{Pauly}%
  \BibitemOpen
  \bibfield  {author} {\bibinfo {author} {\bibfnamefont {H.}~\bibnamefont
  {Pauly}},\ }\href@noop {} {\emph {\bibinfo {title} {{Atom, Molecule and
  Clusterbeams I: Basic Theory, Production and Detection of Thermal Beams}}}}\
  (\bibinfo  {publisher} {Springer-Verlag, Berlin},\ \bibinfo {year}
  {2000})\BibitemShut {NoStop}%
\bibitem [{\citenamefont {Pauly}(2013)}]{pauly2013atom}%
  \BibitemOpen
  \bibfield  {author} {\bibinfo {author} {\bibfnamefont {H.}~\bibnamefont
  {Pauly}},\ }\href@noop {} {\emph {\bibinfo {title} {{Atom, Molecule, and
  Cluster Beams II: Cluster Beams, Fast and Slow Beams, Accessory Equipment and
  Applications}}}},\ Vol.~\bibinfo {volume} {32}\ (\bibinfo  {publisher}
  {Springer Science {\&} Business Media},\ \bibinfo {year} {2013})\BibitemShut
  {NoStop}%
\bibitem [{\citenamefont {Lucas}(2013)}]{lucas2013atomic}%
  \BibitemOpen
  \bibfield  {author} {\bibinfo {author} {\bibfnamefont {C.~B.}\ \bibnamefont
  {Lucas}},\ }\href@noop {} {\emph {\bibinfo {title} {{Atomic and Molecular
  Beams: Production and Collimation}}}}\ (\bibinfo  {publisher} {CRC Press},\
  \bibinfo {year} {2013})\BibitemShut {NoStop}%
\bibitem [{\citenamefont {Morse}(1996)}]{Morse1996}%
  \BibitemOpen
  \bibfield  {author} {\bibinfo {author} {\bibfnamefont {M.~D.}\ \bibnamefont
  {Morse}},\ }in\ \href {\doibase 10.1016/S0076-695X(08)60784-X} {\emph
  {\bibinfo {booktitle} {Exp. Methods Phys. Sci.}}},\ Vol.~\bibinfo {volume}
  {29}\ (\bibinfo {year} {1996})\ pp.\ \bibinfo {pages} {21--47}\BibitemShut
  {NoStop}%
\bibitem [{\citenamefont {Sinha}\ \emph {et~al.}(1973)\citenamefont {Sinha},
  \citenamefont {Schultz},\ and\ \citenamefont {Zare}}]{sinha1973internal}%
  \BibitemOpen
  \bibfield  {author} {\bibinfo {author} {\bibfnamefont {M.~P.}\ \bibnamefont
  {Sinha}}, \bibinfo {author} {\bibfnamefont {A.}~\bibnamefont {Schultz}}, \
  and\ \bibinfo {author} {\bibfnamefont {R.~N.}\ \bibnamefont {Zare}},\
  }\href@noop {} {\bibfield  {journal} {\bibinfo  {journal} {J. Chem. Phys.}\
  }\textbf {\bibinfo {volume} {58}},\ \bibinfo {pages} {549} (\bibinfo {year}
  {1973})}\BibitemShut {NoStop}%
\bibitem [{\citenamefont {McClelland}\ \emph {et~al.}(1979)\citenamefont
  {McClelland}, \citenamefont {Saenger}, \citenamefont {Valentini},\ and\
  \citenamefont {Herschbach}}]{mcclelland1979vibrational}%
  \BibitemOpen
  \bibfield  {author} {\bibinfo {author} {\bibfnamefont {G.~M.}\ \bibnamefont
  {McClelland}}, \bibinfo {author} {\bibfnamefont {K.~L.}\ \bibnamefont
  {Saenger}}, \bibinfo {author} {\bibfnamefont {J.~J.}\ \bibnamefont
  {Valentini}}, \ and\ \bibinfo {author} {\bibfnamefont {D.~R.}\ \bibnamefont
  {Herschbach}},\ }\href@noop {} {\bibfield  {journal} {\bibinfo  {journal} {J.
  Phys. Chem.}\ }\textbf {\bibinfo {volume} {83}},\ \bibinfo {pages} {947}
  (\bibinfo {year} {1979})}\BibitemShut {NoStop}%
\bibitem [{\citenamefont {Zacharias}\ \emph {et~al.}(1984)\citenamefont
  {Zacharias}, \citenamefont {Loy}, \citenamefont {Roland},\ and\ \citenamefont
  {Sudbo}}]{zacharias1984rotational}%
  \BibitemOpen
  \bibfield  {author} {\bibinfo {author} {\bibfnamefont {H.}~\bibnamefont
  {Zacharias}}, \bibinfo {author} {\bibfnamefont {M.~M.~T.}\ \bibnamefont
  {Loy}}, \bibinfo {author} {\bibfnamefont {P.~A.}\ \bibnamefont {Roland}}, \
  and\ \bibinfo {author} {\bibfnamefont {A.~S.}\ \bibnamefont {Sudbo}},\
  }\href@noop {} {\bibfield  {journal} {\bibinfo  {journal} {J. Chem. Phys.}\
  }\textbf {\bibinfo {volume} {81}},\ \bibinfo {pages} {3148} (\bibinfo {year}
  {1984})}\BibitemShut {NoStop}%
\bibitem [{\citenamefont {Carr}\ \emph {et~al.}(2009)\citenamefont {Carr},
  \citenamefont {DeMille}, \citenamefont {Krems},\ and\ \citenamefont
  {Ye}}]{Carr2009}%
  \BibitemOpen
  \bibfield  {author} {\bibinfo {author} {\bibfnamefont {L.~D.}\ \bibnamefont
  {Carr}}, \bibinfo {author} {\bibfnamefont {D.}~\bibnamefont {DeMille}},
  \bibinfo {author} {\bibfnamefont {R.~V.}\ \bibnamefont {Krems}}, \ and\
  \bibinfo {author} {\bibfnamefont {J.}~\bibnamefont {Ye}},\ }\href {\doibase
  10.1088/1367-2630/11/5/055049} {\bibfield  {journal} {\bibinfo  {journal}
  {New J. Phys.}\ }\textbf {\bibinfo {volume} {11}},\ \bibinfo {pages} {055049}
  (\bibinfo {year} {2009})}\BibitemShut {NoStop}%
\bibitem [{\citenamefont {Hudson}\ \emph {et~al.}(2011)\citenamefont {Hudson},
  \citenamefont {Kara}, \citenamefont {Smallman}, \citenamefont {Sauer},
  \citenamefont {Tarbutt},\ and\ \citenamefont {Hinds}}]{Hudson2011}%
  \BibitemOpen
  \bibfield  {author} {\bibinfo {author} {\bibfnamefont {J.~J.}\ \bibnamefont
  {Hudson}}, \bibinfo {author} {\bibfnamefont {D.~M.}\ \bibnamefont {Kara}},
  \bibinfo {author} {\bibfnamefont {I.~J.}\ \bibnamefont {Smallman}}, \bibinfo
  {author} {\bibfnamefont {B.~E.}\ \bibnamefont {Sauer}}, \bibinfo {author}
  {\bibfnamefont {M.~R.}\ \bibnamefont {Tarbutt}}, \ and\ \bibinfo {author}
  {\bibfnamefont {E.~a.}\ \bibnamefont {Hinds}},\ }\href {\doibase
  10.1038/nature10104} {\bibfield  {journal} {\bibinfo  {journal} {Nature}\
  }\textbf {\bibinfo {volume} {473}},\ \bibinfo {pages} {493} (\bibinfo {year}
  {2011})}\BibitemShut {NoStop}%
\bibitem [{\citenamefont {Jin}\ and\ \citenamefont {Ye}(2012)}]{Jin2012}%
  \BibitemOpen
  \bibfield  {author} {\bibinfo {author} {\bibfnamefont {D.~S.}\ \bibnamefont
  {Jin}}\ and\ \bibinfo {author} {\bibfnamefont {J.}~\bibnamefont {Ye}},\
  }\href {\doibase 10.1021/cr300342x} {\bibfield  {journal} {\bibinfo
  {journal} {Chem. Rev.}\ }\textbf {\bibinfo {volume} {112}},\ \bibinfo {pages}
  {4801} (\bibinfo {year} {2012})}\BibitemShut {NoStop}%
\bibitem [{\citenamefont {Koch}(2016)}]{Koch2016}%
  \BibitemOpen
  \bibfield  {author} {\bibinfo {author} {\bibfnamefont {C.~P.}\ \bibnamefont
  {Koch}},\ }\href {\doibase 10.1088/0953-8984/28/21/213001} {\bibfield
  {journal} {\bibinfo  {journal} {J. Phys. Condens. Matter}\ }\textbf {\bibinfo
  {volume} {28}},\ \bibinfo {pages} {213001} (\bibinfo {year}
  {2016})}\BibitemShut {NoStop}%
\bibitem [{\citenamefont {Wall}(2016)}]{Wall2016}%
  \BibitemOpen
  \bibfield  {author} {\bibinfo {author} {\bibfnamefont {T.~E.}\ \bibnamefont
  {Wall}},\ }\href {\doibase 10.1088/0953-4075/49/24/243001} {\bibfield
  {journal} {\bibinfo  {journal} {J. Phys. B At. Mol. Opt. Phys.}\ }\textbf
  {\bibinfo {volume} {49}},\ \bibinfo {pages} {243001} (\bibinfo {year}
  {2016})}\BibitemShut {NoStop}%
\bibitem [{\citenamefont {Shapiro}\ and\ \citenamefont
  {Brumer}(2003)}]{Shapiro2003}%
  \BibitemOpen
  \bibfield  {author} {\bibinfo {author} {\bibfnamefont {M.}~\bibnamefont
  {Shapiro}}\ and\ \bibinfo {author} {\bibfnamefont {P.}~\bibnamefont
  {Brumer}},\ }\href@noop {} {\emph {\bibinfo {title} {{Principles of the
  Quantum Control of Molecular Processes}}}}\ (\bibinfo  {publisher}
  {Wiley-Interscience, Hoboken, NJ},\ \bibinfo {year} {2003})\BibitemShut
  {NoStop}%
\bibitem [{\citenamefont {D'Alessandro}(2007)}]{DAlessandro2007}%
  \BibitemOpen
  \bibfield  {author} {\bibinfo {author} {\bibfnamefont {D.}~\bibnamefont
  {D'Alessandro}},\ }\href@noop {} {\emph {\bibinfo {title} {{Introduction to
  Quantum Control and Dynamics}}}}\ (\bibinfo  {publisher} {Chapman and
  Hall,Boca Raton},\ \bibinfo {year} {2007})\BibitemShut {NoStop}%
\bibitem [{\citenamefont {Qu{\'{e}}m{\'{e}}ner}\ and\ \citenamefont
  {Julienne}(2012)}]{Quemener2012}%
  \BibitemOpen
  \bibfield  {author} {\bibinfo {author} {\bibfnamefont {G.}~\bibnamefont
  {Qu{\'{e}}m{\'{e}}ner}}\ and\ \bibinfo {author} {\bibfnamefont {P.~S.}\
  \bibnamefont {Julienne}},\ }\href {\doibase 10.1021/cr300092g} {\bibfield
  {journal} {\bibinfo  {journal} {Chem. Rev.}\ }\textbf {\bibinfo {volume}
  {112}},\ \bibinfo {pages} {4949} (\bibinfo {year} {2012})}\BibitemShut
  {NoStop}%
\bibitem [{\citenamefont {Jankunas}\ and\ \citenamefont
  {Osterwalder}(2015)}]{jankunas2015cold}%
  \BibitemOpen
  \bibfield  {author} {\bibinfo {author} {\bibfnamefont {J.}~\bibnamefont
  {Jankunas}}\ and\ \bibinfo {author} {\bibfnamefont {A.}~\bibnamefont
  {Osterwalder}},\ }\href@noop {} {\bibfield  {journal} {\bibinfo  {journal}
  {Annu. Rev. Phys. Chem.}\ }\textbf {\bibinfo {volume} {66}},\ \bibinfo
  {pages} {241} (\bibinfo {year} {2015})}\BibitemShut {NoStop}%
\bibitem [{\citenamefont {Lemeshko}\ \emph {et~al.}(2013)\citenamefont
  {Lemeshko}, \citenamefont {Krems}, \citenamefont {Doyle},\ and\ \citenamefont
  {Kais}}]{Lemeshko2013}%
  \BibitemOpen
  \bibfield  {author} {\bibinfo {author} {\bibfnamefont {M.}~\bibnamefont
  {Lemeshko}}, \bibinfo {author} {\bibfnamefont {R.~V.}\ \bibnamefont {Krems}},
  \bibinfo {author} {\bibfnamefont {J.~M.}\ \bibnamefont {Doyle}}, \ and\
  \bibinfo {author} {\bibfnamefont {S.}~\bibnamefont {Kais}},\ }\href {\doibase
  10.1080/00268976.2013.813595} {\bibfield  {journal} {\bibinfo  {journal}
  {Mol. Phys.}\ }\textbf {\bibinfo {volume} {111}},\ \bibinfo {pages} {1648}
  (\bibinfo {year} {2013})},\ \Eprint {http://arxiv.org/abs/arXiv:1306.0912v2}
  {arXiv:arXiv:1306.0912v2} \BibitemShut {NoStop}%
\bibitem [{\citenamefont {Viteau}\ \emph {et~al.}(2008)\citenamefont {Viteau},
  \citenamefont {Chotia}, \citenamefont {Allegrini}, \citenamefont {Bouloufa},
  \citenamefont {Dulieu}, \citenamefont {Comparat},\ and\ \citenamefont
  {Pillet}}]{Viteau2008a}%
  \BibitemOpen
  \bibfield  {author} {\bibinfo {author} {\bibfnamefont {M.}~\bibnamefont
  {Viteau}}, \bibinfo {author} {\bibfnamefont {A.}~\bibnamefont {Chotia}},
  \bibinfo {author} {\bibfnamefont {M.}~\bibnamefont {Allegrini}}, \bibinfo
  {author} {\bibfnamefont {N.}~\bibnamefont {Bouloufa}}, \bibinfo {author}
  {\bibfnamefont {O.}~\bibnamefont {Dulieu}}, \bibinfo {author} {\bibfnamefont
  {D.}~\bibnamefont {Comparat}}, \ and\ \bibinfo {author} {\bibfnamefont
  {P.}~\bibnamefont {Pillet}},\ }\href {\doibase 10.1126/science.1159496}
  {\bibfield  {journal} {\bibinfo  {journal} {Science}\ }\textbf {\bibinfo
  {volume} {321}},\ \bibinfo {pages} {232} (\bibinfo {year}
  {2008})}\BibitemShut {NoStop}%
\bibitem [{\citenamefont {Manai}\ \emph {et~al.}(2012)\citenamefont {Manai},
  \citenamefont {Horchani}, \citenamefont {Lignier}, \citenamefont {Pillet},
  \citenamefont {Comparat}, \citenamefont {Fioretti},\ and\ \citenamefont
  {Allegrini}}]{Manai2012a}%
  \BibitemOpen
  \bibfield  {author} {\bibinfo {author} {\bibfnamefont {I.}~\bibnamefont
  {Manai}}, \bibinfo {author} {\bibfnamefont {R.}~\bibnamefont {Horchani}},
  \bibinfo {author} {\bibfnamefont {H.}~\bibnamefont {Lignier}}, \bibinfo
  {author} {\bibfnamefont {P.}~\bibnamefont {Pillet}}, \bibinfo {author}
  {\bibfnamefont {D.}~\bibnamefont {Comparat}}, \bibinfo {author}
  {\bibfnamefont {A.}~\bibnamefont {Fioretti}}, \ and\ \bibinfo {author}
  {\bibfnamefont {M.}~\bibnamefont {Allegrini}},\ }\href {\doibase
  10.1103/PhysRevLett.109.183001} {\bibfield  {journal} {\bibinfo  {journal}
  {Phys. Rev. Lett.}\ }\textbf {\bibinfo {volume} {109}},\ \bibinfo {pages}
  {183001} (\bibinfo {year} {2012})}\BibitemShut {NoStop}%
\bibitem [{\citenamefont {Gl{\"{o}}ckner}\ \emph {et~al.}(2015)\citenamefont
  {Gl{\"{o}}ckner}, \citenamefont {Prehn}, \citenamefont {Englert},
  \citenamefont {Rempe},\ and\ \citenamefont {Zeppenfeld}}]{Glockner2015}%
  \BibitemOpen
  \bibfield  {author} {\bibinfo {author} {\bibfnamefont {R.}~\bibnamefont
  {Gl{\"{o}}ckner}}, \bibinfo {author} {\bibfnamefont {A.}~\bibnamefont
  {Prehn}}, \bibinfo {author} {\bibfnamefont {B.~G.~U.}\ \bibnamefont
  {Englert}}, \bibinfo {author} {\bibfnamefont {G.}~\bibnamefont {Rempe}}, \
  and\ \bibinfo {author} {\bibfnamefont {M.}~\bibnamefont {Zeppenfeld}},\
  }\href {\doibase 10.1103/PhysRevLett.115.233001} {\bibfield  {journal}
  {\bibinfo  {journal} {Phys. Rev. Lett.}\ }\textbf {\bibinfo {volume} {115}},\
  \bibinfo {pages} {233001} (\bibinfo {year} {2015})},\ \Eprint
  {http://arxiv.org/abs/1511.07360} {arXiv:1511.07360} \BibitemShut {NoStop}%
\bibitem [{\citenamefont {Hamamda}\ \emph {et~al.}(2015)\citenamefont
  {Hamamda}, \citenamefont {Pillet}, \citenamefont {Lignier},\ and\
  \citenamefont {Comparat}}]{Hamamda2015a}%
  \BibitemOpen
  \bibfield  {author} {\bibinfo {author} {\bibfnamefont {M.}~\bibnamefont
  {Hamamda}}, \bibinfo {author} {\bibfnamefont {P.}~\bibnamefont {Pillet}},
  \bibinfo {author} {\bibfnamefont {H.}~\bibnamefont {Lignier}}, \ and\
  \bibinfo {author} {\bibfnamefont {D.}~\bibnamefont {Comparat}},\ }\href
  {\doibase 10.1088/0953-4075/48/18/182001} {\bibfield  {journal} {\bibinfo
  {journal} {J. Phys. B At. Mol. Opt. Phys.}\ }\textbf {\bibinfo {volume}
  {48}},\ \bibinfo {pages} {182001} (\bibinfo {year} {2015})}\BibitemShut
  {NoStop}%
\bibitem [{\citenamefont {Staanum}\ \emph {et~al.}(2010)\citenamefont
  {Staanum}, \citenamefont {H{\o}jbjerre}, \citenamefont {Skyt}, \citenamefont
  {Hansen},\ and\ \citenamefont {Drewsen}}]{Staanum2010}%
  \BibitemOpen
  \bibfield  {author} {\bibinfo {author} {\bibfnamefont {P.~F.}\ \bibnamefont
  {Staanum}}, \bibinfo {author} {\bibfnamefont {K.}~\bibnamefont
  {H{\o}jbjerre}}, \bibinfo {author} {\bibfnamefont {P.~S.}\ \bibnamefont
  {Skyt}}, \bibinfo {author} {\bibfnamefont {A.~K.}\ \bibnamefont {Hansen}}, \
  and\ \bibinfo {author} {\bibfnamefont {M.}~\bibnamefont {Drewsen}},\ }\href
  {\doibase 10.1038/nphys1604} {\bibfield  {journal} {\bibinfo  {journal} {Nat.
  Phys.}\ }\textbf {\bibinfo {volume} {6}},\ \bibinfo {pages} {271} (\bibinfo
  {year} {2010})}\BibitemShut {NoStop}%
\bibitem [{\citenamefont {Schneider}\ \emph {et~al.}(2010)\citenamefont
  {Schneider}, \citenamefont {Roth}, \citenamefont {Duncker}, \citenamefont
  {Ernsting},\ and\ \citenamefont {Schiller}}]{Schneider2010}%
  \BibitemOpen
  \bibfield  {author} {\bibinfo {author} {\bibfnamefont {T.}~\bibnamefont
  {Schneider}}, \bibinfo {author} {\bibfnamefont {B.}~\bibnamefont {Roth}},
  \bibinfo {author} {\bibfnamefont {H.}~\bibnamefont {Duncker}}, \bibinfo
  {author} {\bibfnamefont {I.}~\bibnamefont {Ernsting}}, \ and\ \bibinfo
  {author} {\bibfnamefont {S.}~\bibnamefont {Schiller}},\ }\href {\doibase
  10.1038/nphys1605} {\bibfield  {journal} {\bibinfo  {journal} {Nat. Phys.}\
  }\textbf {\bibinfo {volume} {6}},\ \bibinfo {pages} {275} (\bibinfo {year}
  {2010})}\BibitemShut {NoStop}%
\bibitem [{\citenamefont {Yzombard}\ \emph {et~al.}(2015)\citenamefont
  {Yzombard}, \citenamefont {Hamamda}, \citenamefont {Gerber}, \citenamefont
  {Doser},\ and\ \citenamefont {Comparat}}]{Yzombard2015}%
  \BibitemOpen
  \bibfield  {author} {\bibinfo {author} {\bibfnamefont {P.}~\bibnamefont
  {Yzombard}}, \bibinfo {author} {\bibfnamefont {M.}~\bibnamefont {Hamamda}},
  \bibinfo {author} {\bibfnamefont {S.}~\bibnamefont {Gerber}}, \bibinfo
  {author} {\bibfnamefont {M.}~\bibnamefont {Doser}}, \ and\ \bibinfo {author}
  {\bibfnamefont {D.}~\bibnamefont {Comparat}},\ }\href {\doibase
  10.1103/PhysRevLett.114.213001} {\bibfield  {journal} {\bibinfo  {journal}
  {Phys. Rev. Lett.}\ }\textbf {\bibinfo {volume} {114}},\ \bibinfo {pages}
  {213001} (\bibinfo {year} {2015})}\BibitemShut {NoStop}%
\bibitem [{\citenamefont {Lien}\ \emph {et~al.}(2014)\citenamefont {Lien},
  \citenamefont {Seck}, \citenamefont {Lin}, \citenamefont {Nguyen},
  \citenamefont {Tabor},\ and\ \citenamefont {Odom}}]{Lien2014}%
  \BibitemOpen
  \bibfield  {author} {\bibinfo {author} {\bibfnamefont {C.-Y.}\ \bibnamefont
  {Lien}}, \bibinfo {author} {\bibfnamefont {C.~M.}\ \bibnamefont {Seck}},
  \bibinfo {author} {\bibfnamefont {Y.-W.}\ \bibnamefont {Lin}}, \bibinfo
  {author} {\bibfnamefont {J.~H.~V.}\ \bibnamefont {Nguyen}}, \bibinfo {author}
  {\bibfnamefont {D.~A.}\ \bibnamefont {Tabor}}, \ and\ \bibinfo {author}
  {\bibfnamefont {B.~C.}\ \bibnamefont {Odom}},\ }\href {\doibase
  10.1038/ncomms5783} {\bibfield  {journal} {\bibinfo  {journal} {Nat.
  Commun.}\ }\textbf {\bibinfo {volume} {5}},\ \bibinfo {pages} {4783}
  (\bibinfo {year} {2014})}\BibitemShut {NoStop}%
\bibitem [{\citenamefont {Campbell}\ and\ \citenamefont
  {Doyle}(2009)}]{Campbell2009}%
  \BibitemOpen
  \bibfield  {author} {\bibinfo {author} {\bibfnamefont {W.~C.}\ \bibnamefont
  {Campbell}}\ and\ \bibinfo {author} {\bibfnamefont {J.}~\bibnamefont
  {Doyle}},\ }\href@noop {} {\bibfield  {journal} {\bibinfo  {journal} {Cold
  Mol. Theory, Exp. Appl.}\ ,\ \bibinfo {pages} {473}} (\bibinfo {year}
  {2009})}\BibitemShut {NoStop}%
\bibitem [{\citenamefont {Barry}\ \emph {et~al.}(2011)\citenamefont {Barry},
  \citenamefont {Shuman},\ and\ \citenamefont {DeMille}}]{Barry2011}%
  \BibitemOpen
  \bibfield  {author} {\bibinfo {author} {\bibfnamefont {J.~F.}\ \bibnamefont
  {Barry}}, \bibinfo {author} {\bibfnamefont {E.~S.}\ \bibnamefont {Shuman}}, \
  and\ \bibinfo {author} {\bibfnamefont {D.}~\bibnamefont {DeMille}},\ }\href
  {\doibase 10.1039/c1cp20335e} {\bibfield  {journal} {\bibinfo  {journal}
  {Phys. Chem. Chem. Phys.}\ }\textbf {\bibinfo {volume} {13}},\ \bibinfo
  {pages} {18936} (\bibinfo {year} {2011})},\ \Eprint
  {http://arxiv.org/abs/1101.4229} {arXiv:1101.4229} \BibitemShut {NoStop}%
\bibitem [{\citenamefont {Chen}\ \emph {et~al.}(2016)\citenamefont {Chen},
  \citenamefont {Bu},\ and\ \citenamefont {Yan}}]{Chen2016}%
  \BibitemOpen
  \bibfield  {author} {\bibinfo {author} {\bibfnamefont {T.}~\bibnamefont
  {Chen}}, \bibinfo {author} {\bibfnamefont {W.}~\bibnamefont {Bu}}, \ and\
  \bibinfo {author} {\bibfnamefont {B.}~\bibnamefont {Yan}},\ }\href {\doibase
  10.1103/PhysRevA.94.063415} {\bibfield  {journal} {\bibinfo  {journal} {Phys.
  Rev. A}\ }\textbf {\bibinfo {volume} {94}},\ \bibinfo {pages} {063415}
  (\bibinfo {year} {2016})}\BibitemShut {NoStop}%
\bibitem [{\citenamefont {Herzberg}(1950)}]{Herzberg1950}%
  \BibitemOpen
  \bibfield  {author} {\bibinfo {author} {\bibfnamefont {G.}~\bibnamefont
  {Herzberg}},\ }\href@noop {} {\emph {\bibinfo {title} {{Molecular spectra and
  molecular structure: I. Spectra of diatomic molecules}}}},\ \bibinfo
  {edition} {2nd}\ ed.\ (\bibinfo  {publisher} {Krieger Publishing Company},\
  \bibinfo {address} {Malabar},\ \bibinfo {year} {1950})\BibitemShut {NoStop}%
\bibitem [{\citenamefont {Steimle}\ \emph {et~al.}(2011)\citenamefont
  {Steimle}, \citenamefont {Frey}, \citenamefont {Le}, \citenamefont {DeMille},
  \citenamefont {Rahmlow},\ and\ \citenamefont {Linton}}]{Steimle2011}%
  \BibitemOpen
  \bibfield  {author} {\bibinfo {author} {\bibfnamefont {T.~C.}\ \bibnamefont
  {Steimle}}, \bibinfo {author} {\bibfnamefont {S.}~\bibnamefont {Frey}},
  \bibinfo {author} {\bibfnamefont {A.}~\bibnamefont {Le}}, \bibinfo {author}
  {\bibfnamefont {D.}~\bibnamefont {DeMille}}, \bibinfo {author} {\bibfnamefont
  {D.~A.}\ \bibnamefont {Rahmlow}}, \ and\ \bibinfo {author} {\bibfnamefont
  {C.}~\bibnamefont {Linton}},\ }\href {\doibase 10.1103/PhysRevA.84.012508}
  {\bibfield  {journal} {\bibinfo  {journal} {Phys. Rev. A}\ }\textbf {\bibinfo
  {volume} {84}},\ \bibinfo {pages} {012508} (\bibinfo {year}
  {2011})}\BibitemShut {NoStop}%
\bibitem [{\citenamefont {Effantin}\ \emph {et~al.}(1990)\citenamefont
  {Effantin}, \citenamefont {Bernard}, \citenamefont {D'Incani}, \citenamefont
  {Wannousw}, \citenamefont {Verg{\`{e}}s},\ and\ \citenamefont
  {Barrow}}]{Effantin1990a}%
  \BibitemOpen
  \bibfield  {author} {\bibinfo {author} {\bibfnamefont {C.}~\bibnamefont
  {Effantin}}, \bibinfo {author} {\bibfnamefont {A.}~\bibnamefont {Bernard}},
  \bibinfo {author} {\bibfnamefont {J.}~\bibnamefont {D'Incani}}, \bibinfo
  {author} {\bibfnamefont {G.}~\bibnamefont {Wannousw}}, \bibinfo {author}
  {\bibfnamefont {J.}~\bibnamefont {Verg{\`{e}}s}}, \ and\ \bibinfo {author}
  {\bibfnamefont {R.~F.}\ \bibnamefont {Barrow}},\ }\href {\doibase
  10.1080/00268979000101311} {\bibfield  {journal} {\bibinfo  {journal} {Mol.
  Phys.}\ }\textbf {\bibinfo {volume} {70}},\ \bibinfo {pages} {735} (\bibinfo
  {year} {1990})}\BibitemShut {NoStop}%
\bibitem [{\citenamefont {Western}(2017)}]{Western2017}%
  \BibitemOpen
  \bibfield  {author} {\bibinfo {author} {\bibfnamefont {C.~M.}\ \bibnamefont
  {Western}},\ }\href {\doibase 10.1016/j.jqsrt.2016.04.010} {\bibfield
  {journal} {\bibinfo  {journal} {J. Quant. Spectrosc. Radiat. Transf.}\
  }\textbf {\bibinfo {volume} {186}},\ \bibinfo {pages} {221} (\bibinfo {year}
  {2017})}\BibitemShut {NoStop}%
\end{thebibliography}

\end{document}